\documentclass{article}

\usepackage{arxiv}

\usepackage[utf8]{inputenc} 
\usepackage[T1]{fontenc}    
\usepackage{hyperref}       
\usepackage{url}            
\usepackage{booktabs}       
\usepackage{amsfonts}       
\usepackage{nicefrac}       
\usepackage{microtype}      
\usepackage{lipsum}		
\usepackage{graphicx}
\usepackage{natbib}
\usepackage{doi}

\title{Similarities in Massive Separation Across Reynolds Numbers for Swept and Tapered Finite Span Wings}

\author{ \href{https://orcid.org/0000-0002-9681-7087}{\includegraphics[scale=0.06]{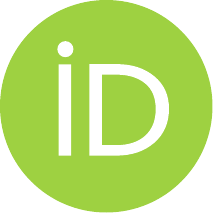}\hspace{1mm}Jacob M. Neal} \\
	Department of Mechanical, \\
        Aeronautical, and Nuclear Engineering\\
	Rensselaer Polytechnic Institute\\
	Troy, NY 12180\\
	\And
	\href{https://orcid.org/0000-0002-8268-9088}{\includegraphics[scale=0.06]{orcid.pdf}\hspace{1mm}Anton Burtsev} \\
	Department of Mechanical\\
 and Aerospace Engineering\\
        University of Liverpool\\
        Liverpool L69 3GH, UK\\
 	\And
	\href{https://orcid.org/0000-0002-8613-9593}{\includegraphics[scale=0.06]{orcid.pdf}\hspace{1mm}Jean H\'{e}lder Marques Ribeiro} \\
	Department of Mechanical \\
        and Aerospace Engineering\\
        University of California \\
	Los Angeles, CA\\
 	\And
	\href{https://orcid.org/0000-0002-3762-8075}{\includegraphics[scale=0.06]{orcid.pdf}\hspace{1mm}Kunihiko Taira} \\
	Department of Mechanical \\
        and Aerospace Engineering\\
        University of California \\
	Los Angeles, CA\\
 	\And
	\href{https://orcid.org/0000-0002-7720-3434}{\includegraphics[scale=0.06]{orcid.pdf}\hspace{1mm}Vassilios Theofilis} \\
	Department of Mechanical\\
    and Aerospace Engineering\\
        University of Liverpool\\
        Liverpool L69 3GH, UK\\
 	\And
	\href{https://orcid.org/0000-0003-1366-0406}{\includegraphics[scale=0.06]{orcid.pdf}\hspace{1mm}Michael Amitay} \\
	Department of Mechanical, \\
        Aeronautical, and Nuclear Engineering\\
	Rensselaer Polytechnic Institute\\
	Troy, NY 12180\\
}

\renewcommand{\shorttitle}{\textit{arXiv} Template}

\hypersetup{
pdftitle={A template for the arxiv style},
pdfsubject={q-bio.NC, q-bio.QM},
pdfauthor={David S.~Hippocampus, Elias D.~Striatum},
pdfkeywords={First keyword, Second keyword, More},
}

\begin{document}
\maketitle

\begin{abstract}
Experimental investigations were performed to elucidate the features of flow fields occurring over cantilevered finite-aspect ratio NACA 0015 wings at high angles of attack with various sweep angles and taper ratios. Volumetric Stereoscopic Particle Image Velocimetry experiments were performed at mean chord based Reynolds number of $247,500$ in a wind tunnel and $600$ in a water tunnel. Direct Numerical Simulations (DNS) of the water tunnel test section, including the cantilevered model, were also performed at the lower Reynolds number. The low Reynolds number experiments, low Reynolds number simulations, and high Reynolds number experiments all showed that sweeping the leading edge back shifted the largest portion of the reversed flow towards the wingtip while sweeping the trailing edge forward shifted the reversed flow towards the wing root. A detailed parametric sweep of planform geometry systematically varied the leading and trailing edge sweep angles and taper ratios of the finite wings.  It was found that the large scale vortical structures resulting from varying these parameters at the two Reynolds numbers share surprisingly many three-dimensional topological features, despite the orders of magnitude different Reynolds numbers.
\end{abstract}

\section{Introduction}
\label{sec:intro}
Flow separation over wings and control surfaces results in substantial decrement of aircraft performance.  Stall on finite span wings can lead to complex three-dimensional flow structures which are highly dependent on planform geometry.  While comprehensive studies of the effects of planform on 3-D separated wakes are limited, recent computational efforts have elucidated the wake dynamics and stability of flows over finite-span wings at very low Reynolds numbers $\mathcal{O}(10^2)$ \citep{Taira2009,He2017,Burtsev2022,ribeiro_yeh_taira_2023}.  

Some of the studies of 3-D flow separation over wings have focused on unswept planforms, where the dominant 3-D structure is the ``owl eye" or ``mushroom-shaped" stall-cell pattern \citep{bippes_n_turk,WEIHS1983,Boiko1996,DellOrso2016a,DellOrso2018,Neal2023}.  Stall cells are made of two counter-rotating surface normal vortices which arise from an interaction of midspan reversed flow and attached flow at the wingtip and root.   Stall cells form in shallow stall \citep{DellOrso2018} and deep stall \citep{Yon1998} over thick airfoils experiencing trailing edge separation \citep{McCullough1951,Broeren2001}.  
The flow field unsteadiness associated with the stall cell has been documented \citep{Manolesos2014,Neal2023}, showing the wake structure and location of the peak Reynolds stresses caused by the shedding of quasi-spanwise vortices from the midspan.  \citet{zhang2020formation} found that unswept wings exhibited midspan separation for the formation of surface foci indicative of the stall cell pattern at $Re_{\bar{c}} = 400$.    On a rectangular, three-dimensional flat plate wing, \citet{Taira2009} compared their numerical results of 3-D separation in impulsively started flow at $Re_{\bar{c}}$ = 500 to the experimental flow visualization study done by \citet{freymuth1987further} at $Re_{\bar{c}}$ = 5200, noting that the large-scale motions of the vortices were analogous in both cases.  It would follow that some large-scale topological patterns can appear across low to high Reynolds numbers. 

Although sweptback planforms are very common in wings and empennages, limited studies exist on the 3-D flow field of a sweptback wing at high angles of attack.  \citet{Black1956} documented the formation of a so-called ``ram's horn" vortex at $Re_{\bar{c}}=500,000$ on the suction surface of a finite span swept and tapered wing, which resulted in a single surface focus near the wing root.  The tendency for this type of vortex to form at different wing configurations was explored extensively by \citet{Poll1986}.  Surface flow topology and aerodynamic loads for sweptback NACA 0012 wings were documented by \citet{Yen2009} at a range of sweep angles and angles of attack.   \citet{Yen2011} and \citet{neal_hire_swept} concluded that stall occurs when the laminar separation bubble bursts over a sweptback wing.  The tendency for sweptback wings is to stall near the wingtip, resulting in a large scale spanwise rotation over the leading edge that grows and morphs into streamwise rotation into the wake.  This is true for low Reynolds numbers as well. \citet{zhang2020laminar} and \citet{Burtsev2022} noted the formation of the ram's horn reminiscent root-to-tip vortical structures over sweptback wings at high angles of attack and $Re_{\bar{c}}= 400$.  

Forward swept and tapered wings were studied by \citet{Breitsamter2001} at $Re_{\bar{c}}=460,000$, who found that at high angles of attack a vortex would form with a similar shape to the ram's horn seen over sweptback wings but opposite orientation, extending from the tip towards the root. \citet{Zhang2022} made a similar discovery computationally at $Re_{\bar{c}} = 400$. Some studies have been performed on the tip vortex \citep{Skinner2020} and surface topology \citep{zhang_jaworski_mcparlin_turner_2019}, and preliminary studies on wakes \citep{Ribeiro:AIAA23,Neal:AIAA23,Burtsev:AIAA23} of swept and tapered wings, but further analysis is required to fully understand the effects of sweep and taper on the large-scale flow structures.  

The goal of the present work is twofold: to experimentally obtain and explore the volumetric structure of 3-D separation over swept and tapered wings between high and low Reynolds number flows, and to compare the low Reynolds number experimental results to low Reynolds number DNS results.  A key benefit in comparing observations obtained from experiments and from DNS is that, should parity of result be reached between the methods, further post-processing of modal and non-modal stability analysis can be performed on the DNS data, taking advantage of the high spatial resolution and instantaneous volumetric flow fields that the experiments lack.  The paper is organized as follows: section \ref{sec:expsetup} describes the experimental and numerical setups and wing geometries considered; section \ref{sec:results} presents the results obtained from wind tunnel experiments, water tunnel experiments, and DNS; and the concluding statements are in section \ref{sec:Conclusion}.

\subsection{Wing Geometry}
\label{sec:models}

\begin{figure}
    \centering
    \includegraphics[width = 0.6\textwidth]{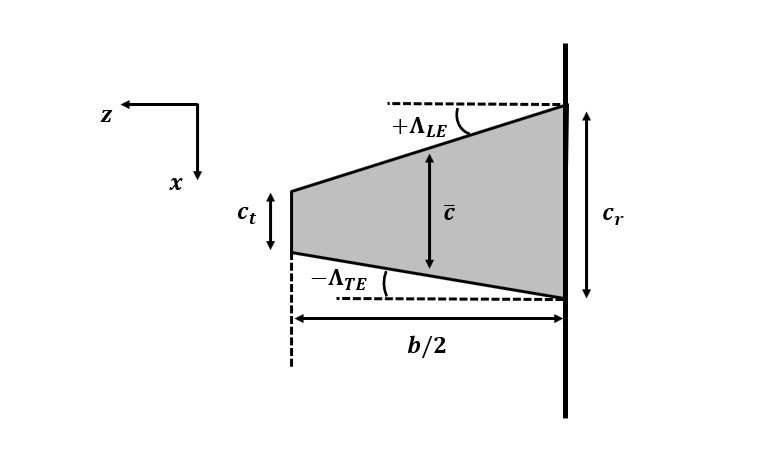}
    \caption{Schematic and definition of parameters for the swept and tapered cantilevered wings all with semi-aspect ratio of 2 that were tested experimentally and numerically. }
    \label{models}
\end{figure}

The parameters of the swept and tapered cantilevered wings examined in this study are displayed in figure \ref{models}.  The $x$, $y$, and $z$ axes are aligned with the freestream, vertical, and spanwise directions, respectively.  All the wings had the NACA 0015 airfoil and a semi-aspect ratio of $sAR = 2$, which is the ratio of the half-span, $b/2$, to the mean chord, $\bar{c}$.  Three taper ratios of $\lambda = c_{t}/c_{r} = 1$, $0.5$,and $0.269$ were explored, where $c_{t}$ and $c_{r}$ are the tip and root chords respectively, and three configurations of sweep angles were studied for each taper ratio giving a total of nine models, as listed in table \ref{tabmodels}.   Geometrically identical planforms were made for wind and water tunnel tests, where the wind tunnel models had a mean chord of $\bar{c}=127\;mm$ and the water tunnel models had a mean chord of $\bar{c}=38.1\;mm$.  

\begin{table}
  \begin{center}
  \begin{tabular}{c c c c}
       $\lambda = c_t/c_r$  &  & $(\Lambda_{LE},\Lambda_{c/4},\Lambda_{TE})$ &  \\ \hline
       1 & $(0^\circ,0^\circ,0^\circ)$ & $(18.4^\circ,18.4^\circ,18.4^\circ)$& $(30^\circ,30^\circ,30^\circ)$\\
       0.5 & $(0^\circ,-4.8^\circ,-18.4^\circ)$ & $(18.4^\circ,14^\circ,0^\circ)$& $(22.6^\circ,18.4^\circ,4.7^\circ)$\\
       0.269 & $(0^\circ,-8.2^\circ,-30^\circ)$ & $(30^\circ,23.4^\circ,0^\circ)$& $(35.8^\circ,30^\circ,8.3^\circ)$\\
  \end{tabular}
  \caption{Taper ratio $\lambda$ and sweep angles of the leading edge, quarter chord, and trailing edge for each of the 9 wing models.}
  \label{tabmodels}
  \end{center}
\end{table}

\subsection{Experimental Facilities}

Experiments were performed at Rensselaer Polytechnic Institute’s Center for Flow Physics and Control (\textit{CeFPaC}) to analyze the effects of sweep and taper on the flow field over cantilevered wings and in their wake.  Oil flow visualizations and Stereoscopic Particle Image Velocimetry (SPIV) measurements were performed in \textit{CeFPaC}'s subsonic open return wind tunnel to qualify and quantify the flow fields at a moderate Reynolds number.  The wind tunnel has a test section that is $5\; m$ long with a $0.8\; m \times 0.8\; m$ cross section and a maximum speed of $50 \; m/s$. Upstream of the test section, flow passes through a flow conditioning unit that is composed of a honeycomb and screens, resulting in a freestream turbulence intensity of less than $0.2 \%$.  The wing model was mounted at mid-height within the wind tunnel test section to minimize tunnel wall effects. Additionally, a fence starts a wall boundary layer $250\;mm$ upstream of the mounted model.  Oil flow visualization and SPIV measurements were performed following the approach of \citet{neal_hire_swept,Neal2023}.

The flow fields at the very low Reynolds number were quantified using SPIV in $CeFPaC$'s ultra-low speed water tunnel.  The closed-return open test section water tunnel had a cross section of $500 \; mm \times 500 \; mm$ with a test section length of $1500 \; mm$. Upstream of the test section and contraction, two perforated plates, turning vanes, a coarse mesh screen, a fine honeycomb screen, and a bug screen were placed to provide for low turbulence intensity and uniform flow into the test section. The area contraction ratio was $6:1$. An AMT heavy-duty centrifugal pump with a maximum power of $7.5 \; hp$ was used with a variable frequency driver to control the speed of the tunnel. The free stream velocity was monitored using an Omega electromagnetic flow meter and verified using SPIV in the free stream.  The wing models were mounted on a fence piece for the SPIV measurement.  SPIV data collection was performed following the approach of \citet{hayoss19,Neal:AIAA23}.

\section{Setup and Procedure}
\label{sec:expsetup}

\subsection{Numerical setup}
\label{sec:num_setup}

The 3-D flow fields over the stalled swept and tapered cantilevered wings were also investigated with direct numerical simulation (DNS) of the entire wind tunnel test section, including the cantilevered model, at geometric and flow parameters identical to those used in the experiment, shown in section \ref{sec:models}, in order to make direct comparisons between experiment and DNS at the low Reynolds number.  The same wing geometries, as discussed in section~\ref{sec:models} were used, and the simulations were carried out at $Re_{\bar{c}}=600$ to match the water tunnel results. The fence present in the experimental setup is taken into account by introducing a no-slip boundary condition at the root of the wing. For more details on the numerical setup, see \citet{Burtsev:AIAA23}.  The boundary layer profiles developing on the splitter plate at $(x,y)/\bar{c}= (2.6, 0.75) $ from the water tunnel SPIV and the DNS were verified to show good agreement (not included here for brevity).  It should be noted that the boundary layer is very thick at this low Reynolds number, and extends over a quarter of the wing span.

\section{Results}
\label{sec:results}

\begin{figure}
    \centering
    \includegraphics[width = \textwidth]{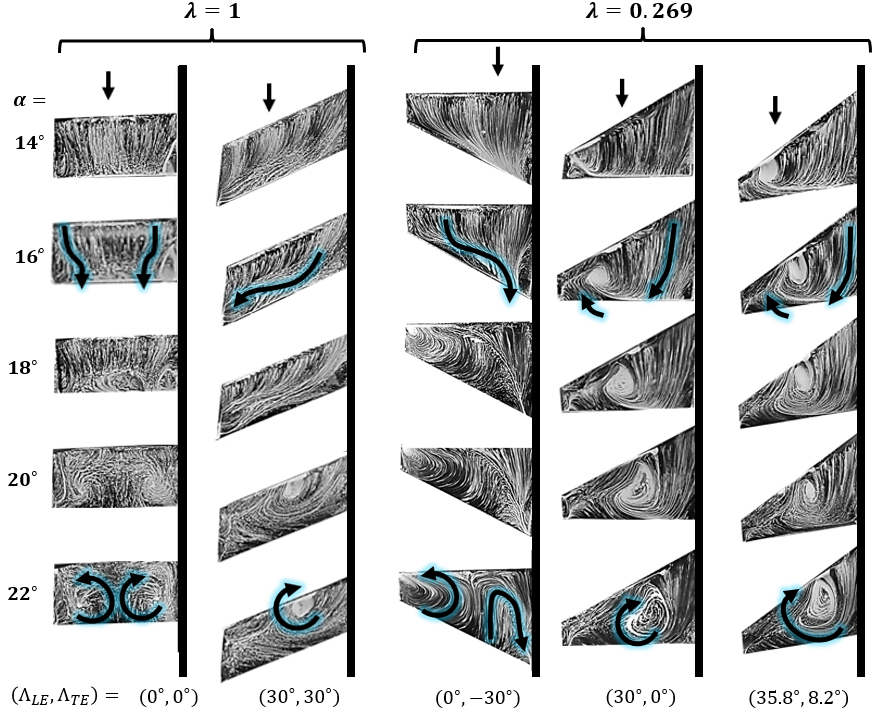}
    \caption{Surface oil flow visualization for $sAR = 2$ cantilevered wings (the wall is on the right side in each image) at $Re_{\bar{c}} = 247,500$, $\alpha = 14^\circ, 16^\circ, 18^\circ, 20^\circ, 22^\circ$, and $\lambda=$1 and 0.269.  Blue arrows denote the flow direction tangent to the wing surface.}
    \label{ofv_all}
\end{figure}

The flow fields over swept and tapered wings at high angles of attack were systematically investigated with a series of wind and water tunnel experiments at $Re_{\bar{c}} = 247,500$ and $Re_{\bar{c}} = 600$, respectively.  In the wind tunnel experiments, the near surface flow topology was first qualitatively explored using the oil flow visualization technique, followed by SPIV measurements of the volumetric mean flow fields over the suction side of the wings and in their wake.  In the water tunnel, the volumetric mean flow fields were measured using SPIV to show that the large scale vortical structures form at both Reynolds numbers.  The water tunnel results at $Re_{\bar{c}}=600$ are also compared to DNS results at the same geometric conditions and Reynolds number.   The DNS results resolve the volumetric mean flow field over the wings at the lower Reynolds number.  

The oil flow visualization technique can give a sense of the effect of sweep and taper on the 3-D separation patterns for the different planforms.  As the goal of this study is to illustrate the similarities in flow structures over swept and tapered wings at very low and moderately high Reynolds numbers, the first step was to identify which flow structures exist for each planform.  The oil flow visualization results are presented in figure \ref{ofv_all} showing the effect of the sweep angle and taper ratio on the surface topology at $Re_{\bar{c}}=247,500$ and several angles of attack.  For all cases, the models are cantilevered from the wall on the right side of each image.  For the unswept untapered wing (left column), the separation begins at around $\alpha = 18^\circ$, where reverse flow is seen at the midspan with two counter-rotating surface foci (marked with the blue arrows at $\alpha = 22^\circ$), which is indicative of a ``mushroom-shaped" stall-cell-type separation.  The surface topology on the sweptback untapered wing (second column from the left) exhibits a region of reversed flow at a similar angle of attack as the unswept wing.  However, in this case the flow field leaves the imprint of a single surface focus rotating in the same direction as the inboard focus in the pair seen on the unswept wing.  This indicates the ram's horn type separation as identified by \citet{Black1956}.  The tapered wing with a forward swept trailing edge $(\Lambda_{LE},\Lambda_{TE})=(0^\circ,-30^\circ)$ (middle column) exhibits the formation of reversed flow near the midspan at around $\alpha = 16^\circ$.  This reversed flow region grows with angle of attack, leading to the formation of a single surface spiral near the wingtip, indicating the formation of an inverted ram's horn vortex similar to the one witnessed on a forward swept wing by \citet{Breitsamter2001}.  For the sweptback tapered models with $(\Lambda_{LE},\Lambda_{TE})=(30^\circ,0^\circ)$ and $(\Lambda_{LE},\Lambda_{TE})=(35.8^\circ,8.2^\circ)$ (two rightmost columns), a small reversed flow region forms near the wingtip at the lower angles of attack, leading to the formation of a single inboard surface spiral.  As angle of attack increases, the reversed flow region grows due to an increasing adverse pressure gradient, thus shifting the surface spiral towards the root.  This is the same behavior of the ``spiral vortex" that \citet{Poll1986} observed with surface visualizations over swept wings.

\begin{figure}
    \centering
    \includegraphics[width = 0.85\textwidth]{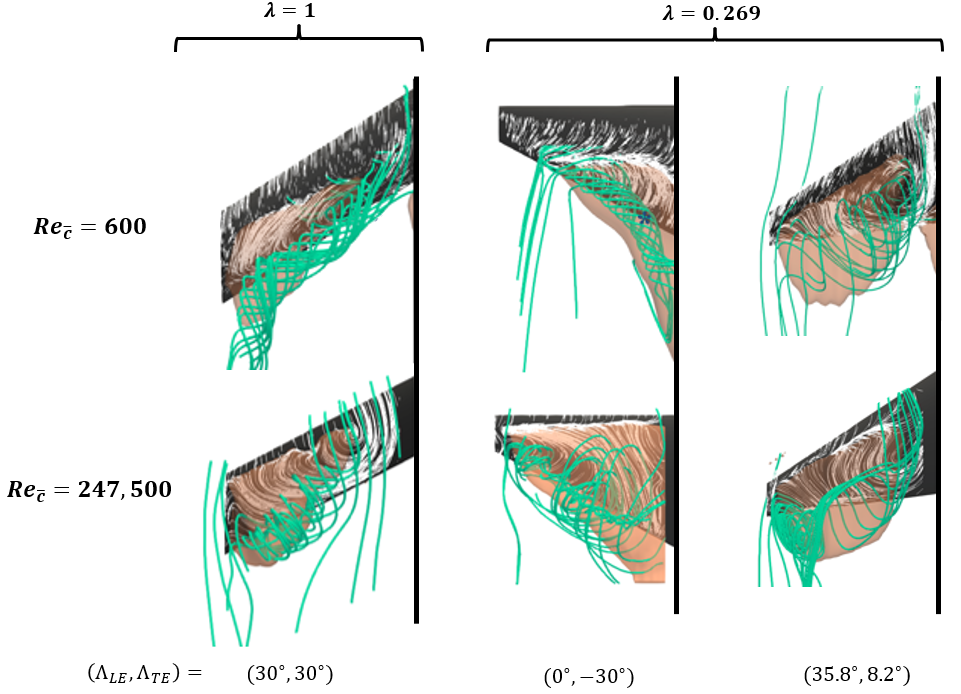}
    \caption{Iso-surface of the reversed flow ($\overline{U}=0$, peach), and 3-D streamlines calculated near the surface (white) and over the wing (green) measured with SPIV at $\alpha = 22^\circ$.  Flow is top to bottom.}
    \label{sepbub}
\end{figure}

The high Reynolds number oil flow visualization shows the effect of sweep and taper on the separated flow.  To quantify these effects, the flow field over the wings was obtained using SPIV in very low Reynolds number water tunnel experiments and moderate Reynolds number wind tunnel experiments.  The global flow structures in the volumetric mean flow fields for representative swept and tapered models at $Re_{\bar{c}}=600$ and $Re_{\bar{c}}=247,500$ are displayed in figure \ref{sepbub}.  The water tunnel results at $Re_{\bar{c}}= 600$ are on the top row, while the wind tunnel results at $Re_{\bar{c}}= 247,500$ are on the bottom row.  For each model, the key aspects of the flow fields are visualized to understand the effect of sweep and taper on the large-scale structures.  The streamlines calculated near the suction surface are colored white, the separation region (3-D contour of $\overline{U}=0$) is colored peach, and select 3-D streamlines calculated through the separation region are colored green.  For the untapered sweptback wing (left column), the near-surface streamlines trace a single surface focus near the leading edge between the reversed flow near the tip and the attached flow near the root, matching the topological pattern seen in the surface visualizations.  The separation region extends from this focus and grows towards the tip.  The 3-D streamlines trace a ram's horn vortex that emanates from the surface focus, extends towards the tip and bends into the streamwise direction in the wake.  

For the tapered TE swept forward wing (middle column), the 3-D streamlines trace a vortex that starts from a surface focus near the tip and extend towards the root (the inverted ram's horn vortex).  The separation region extends from this focus and grows towards the root due to the inboard spanwise velocity.  For the sweptback tapered model with $(\Lambda_{LE},\Lambda_{TE})=(35.8^\circ,8.2^\circ)$ (right column), the separation region is centered near the midspan.  The 3-D streamlines follow a spiraled path from a single surface focus near the root towards the tip.  The flow field for this model more closely resembles the characteristic ram's horn vortex.  The shift of the separation region with sweep in taper is analogous for the low and moderate Reynolds number cases; however, the separation location and overall size of the separation region is different for each regime.  Clearly, the effect of viscosity will be more pronounced  at the low Reynolds number and thus retain attached flow further along the wing, as is seen in the present results.  That said, for all models, the dominant flow features are commonly observed for $Re_{\bar{c}} = 600$ and $247,500$.

\begin{figure}
    \centering
    \includegraphics[width = \textwidth]{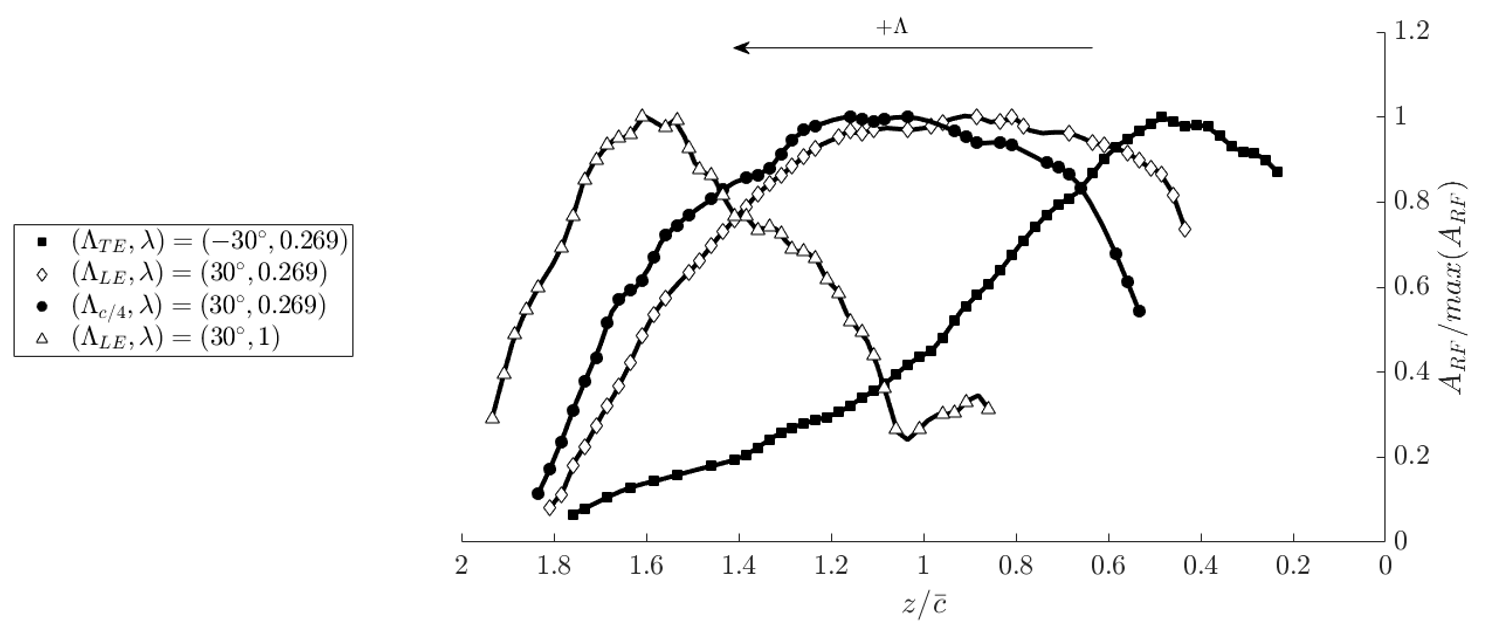}
    \caption{Spanwise distribution of the normalized reversed flow area measured with SPIV at $\alpha = 22^\circ$ and $Re_{\bar{c}} = 247,500$.}
    \label{RF_hi}
\end{figure}

The snapshots of the global flow structures have qualitatively indicated that both the location and severity of the separation are strongly influenced by sweep and taper.  To quantify this effect, the size and extent of the reversed flow were obtained from the volumetric flow fields.  The area of the reversed flow $A_{RF}$ is calculated in each $x-y$ plane as the region enclosing $\bar{U}<0$ at each spanwise location $z/\bar{c}$ and is normalized by peak reversed flow area $\max(A_{RF})$.  The variation of normalized $A_{RF}$ along the wings span for the $Re_{\bar{c}}=247,500$ cases are shown in figure \ref{RF_hi}.  For the wing with a forward swept trailing edge and unswept leading edge, $(\Lambda_{TE},\lambda)=(-30^\circ,0.269)$, the reversed flow is the largest near the root at $z/\bar{c}=0.5$ due to the inverted ram's horn vortex, as was seen in the flow visualization and the SPIV global flow field. For the sweptback and tapered wing with $(\Lambda_{LE},\lambda)=(30^\circ,0.269)$ and $(\Lambda_{c/4},\lambda)=(30^\circ,0.269)$, the reversed flow distribution peaks at the midspan, corresponding to the oil flow visualization seen in figure \ref{ofv_all}.  Note that the $(\Lambda_{c/4},\lambda)=(30^\circ,0.269)$ wing, the leading edge and trailing edge are sweptback slightly further than the $(\Lambda_{LE},\lambda)=(30^\circ,0.269)$ wing, leading to a slight shift of the $A_{RF}$ peak towards the tip.  For the untapered swept wing with $\Lambda = 30^\circ$, the reversed flow distribution exhibits a peak near the wingtip at $z/\bar{c}=1.6$ due to the presence of the ram's horn vortex.  There is a clear pattern of a shifting of the reversed flow peak towards the wingtip with increase in $\Lambda$ and towards the root with decrease in $\Lambda$ for $Re_{\bar{c}}=247,500$.

\begin{figure}
    \centering
    \includegraphics[width = \textwidth]{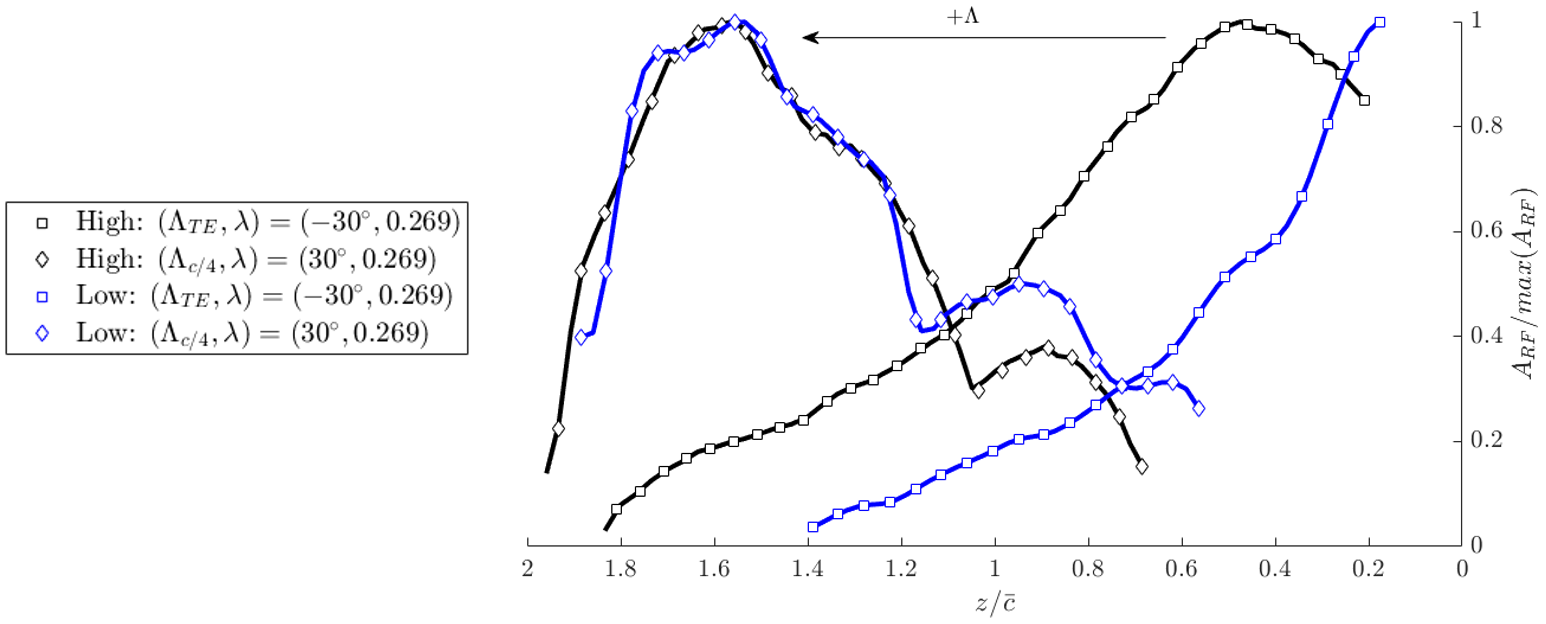}
    \caption{Comparison of the spanwise distribution of the normalized reversed flow area at $Re_{\bar{c}} = 600$ (blue solid symbols) and $Re_{\bar{c}} = 247,500$ (open black symbols) for selected cases at $\alpha = 22^\circ$.}
    \label{RF_hi_v_lo}
\end{figure}

Here, the key comparison is made: the results presented for $Re_{\bar{c}} = 247,500$ are compared with results obtained at $Re_{\bar{c}} = 600$.  The comparison of the spanwise distributions of the reversed flow area for geometrically identical wings at the above mentioned Reynolds numbers is shown in figure \ref{RF_hi_v_lo}.  Even at these vastly different Reynolds numbers, the trend of shifting the peak reversed flow region towards the wingtip with increased $\Lambda$ is captured.  It should be noted that the agreement is far better for the case with the sweptback quarter chord line ($\Lambda_{c/4}=30^\circ$) than for the case of the tapered wing with the forward swept trailing edge ($\Lambda_{TE}=-30$).  In fact, the $A_{RF}$ peaks for the $\Lambda_{c/4}=30^\circ$ model lie on top of one another for both Reynolds numbers, while the $A_{RF}$ peaks of $\Lambda_{TE}=-30$ are separated by about $24\%$ of the span. This is likely caused by the difference in the boundary layer thickness for these cases, where the boundary layer is only $\approx 1\%$ of the wingspan for the moderate Reynolds number case and nearly $25\%$ of the wingspan for the low Reynolds number case.  This encourages the reattachment of the flow near the wall for the moderate Reynolds number case whereas the reversed flow region is increased near the wall for the low Reynolds number case due to much larger horseshoe vortex.  For the wings with the sweptback quarter chord line, the separation occurs further from the wall, and the reduced effect of boundary layer promotes the very good agreement of the normalized of the $A_{RF}$ that is seen at both Reynolds numbers.

\begin{figure}
    \centering
    \includegraphics[width = \textwidth]{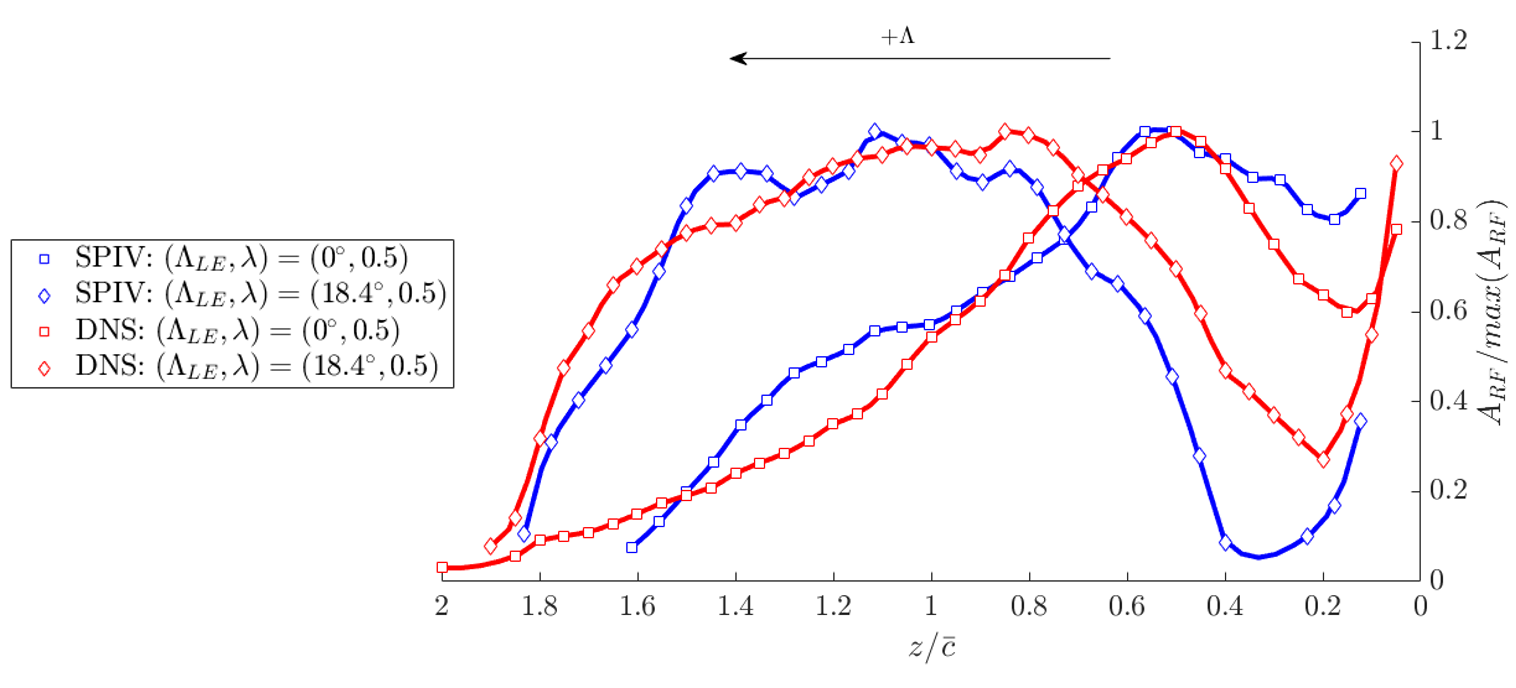}
    \caption{Spanwise distribution of the normalized reversed flow area, calculated with DNS (red solid symbols) and SPIV (black open symbols), at $\alpha = 22^\circ$ and $Re_{\bar{c}} = 600$.  }
    \label{RF_dns_exp}
\end{figure}

Now that the agreement between the low and high Reynolds number experiments has been established, the trends between the low Reynolds number experiments and the low Reynolds number DNS are verified to share the same features.  The spanwise distributions of the normalized reversed flow area from the DNS and the SPIV are presented in figure \ref{RF_dns_exp}.  The models with $(\Lambda_{LE},\lambda)=(0^\circ,0.5)$ and $(\Lambda_{LE},\lambda)=(18.4^\circ,0.5)$ are compared for the SPIV results (open black symbols) and the results obtained from the DNS (solid red symbols).  Both SPIV and DNS results agree well on the trend that increasing $\Lambda_{LE}$ results in a shift of the peak of the reversed flow area towards the wing tip, while sweeping the TE forward shifts the peak separation area towards the wing root. 

The global flow structures of 3-D separation over finite wings are similarly affected by sweep and taper in the present high Reynolds number experiments, low Reynolds number experiments, and low Reynolds number DNS.  From the experiments, the inverted rams horn resulting from forward swept trailing edge caused the shift of the reversed flow towards the root, while the sweptback leading edge causing the formation of the classical ram's horn leading to a shift of the reversed flow towards the tip.  The DNS revealed analogous behaviour as the peak reversed flow area shifted outboard with increasing sweptback angle.  The overall conclusion drawn from the features studied here have provided a fundamental basis to study large-scale separated flow structures over wings at much higher Reynolds numbers.

\section{Conclusion}
\label{sec:Conclusion}

The structure of 3-D separation over swept and tapered cantilevered wings was explored using coupled wind tunnel experiments, water tunnel experiments and DNS.  Untapered unswept wings experience the stall cell type separation while untapered sweptback wings experience the root-to-tip ram's horn type separation pattern.  Wings that are tapered with sweptback leading and trailing edges experience a similar tip stall, which causes the formation of a surface spiral with a similar sense of recirculation as the ram's horn.  Wings with forward swept trailing edge a tip to the root inverted ram's horn vortex.  Qualitatively analogous large-scale flow structures were measured experimentally at $Re_{\bar{c}} = 600$ and at $Re_{\bar{c}} = 247,500$.  The formation of these structures at both Reynolds numbers was linked to a shift in the maximum reversed flow area towards the wingtip with increasing backwards sweep angle, while quantitatively identical results as those at $Re_{\bar{c}} = 600$ were also found in the DNS.  The results herein provide a fundamental understanding of the effect of sweep and taper on separated flow over finite wings.

\section{Acknowledgements}

The authors thank Pam Pulla, Jacob Charest, and Ben Grinberg for aid in the design of the experimental models.  The authors also thank Jack Ottinger, Brandon Gares, and Dr. Shelby Hayostek for their help in collecting the SPIV data.  

\section{Funding}

This work was supported by the Air Force Office of Scientific Research (grant number FA9550-21-1-0174) monitored by Dr. Gregg Abate.

\section{Declaration of interests}
The authors report no conflict of interest.

\bibliographystyle{unsrtnat}
\bibliography{references}

\begin{thebibliography}{30}
\providecommand{\natexlab}[1]{#1}
\providecommand{\url}[1]{\texttt{#1}}
\expandafter\ifx\csname urlstyle\endcsname\relax
  \providecommand{\doi}[1]{doi: #1}\else
  \providecommand{\doi}{doi: \begingroup \urlstyle{rm}\Url}\fi

\bibitem[Taira and Colonius(2009)]{Taira2009}
K.~Taira and T.~Colonius.
\newblock {Three-dimensional flows around low-aspect-ratio flat-plate wings at
  low Reynolds numbers}.
\newblock \emph{Journal of Fluid Mechanics}, 623:\penalty0 187--207, 2009.
\newblock ISSN 00221120.
\newblock \doi{10.1017/S0022112008005314}.

\bibitem[He et~al.(2017)He, Gioria, P{\'{e}}rez, and Theofilis]{He2017}
W.~He, R.~S. Gioria, J.~M. P{\'{e}}rez, and V.~Theofilis.
\newblock {Linear instability of low Reynolds number massively separated flow
  around three NACA airfoils}.
\newblock \emph{Journal of Fluid Mechanics}, 811:\penalty0 701--741, 2017.
\newblock ISSN 14697645.
\newblock \doi{10.1017/jfm.2016.778}.

\bibitem[Burtsev et~al.(2022)Burtsev, He, Zhang, Theofilis, Taira, and
  Amitay]{Burtsev2022}
A.~Burtsev, W.~He, K.~Zhang, Va. Theofilis, K.~Taira, and M.~Amitay.
\newblock {Linear modal instabilities around post-stall swept finite wings at
  low Reynolds numbers}.
\newblock \emph{Journal of Fluid Mechanics}, 944:\penalty0 1--28, 2022.
\newblock ISSN 0022-1120.
\newblock \doi{10.1017/jfm.2022.420}.

\bibitem[Ribeiro et~al.(2023{\natexlab{a}})Ribeiro, Yeh, and
  Taira]{ribeiro_yeh_taira_2023}
J.H.~Marques Ribeiro, Chi-An Yeh, and Kunihiko Taira.
\newblock Triglobal resolvent analysis of swept-wing wakes.
\newblock \emph{Journal of Fluid Mechanics}, 954:\penalty0 A42,
  2023{\natexlab{a}}.
\newblock \doi{10.1017/jfm.2022.1033}.

\bibitem[Bippes and Turk(1980)]{bippes_n_turk}
H.~Bippes and M.~Turk.
\newblock Windkanalmessungen in einem rechteckflügel bei anliegender und
  abgelöster strömung.
\newblock Technical report, DFVLR, DFVLR-IB 251-80, 1980.

\bibitem[Weihs and Katz(1983)]{WEIHS1983}
D.~Weihs and J.~Katz.
\newblock {Cellular patterns in poststall flow over unswept wings}.
\newblock \emph{AIAA Journal}, 21\penalty0 (12):\penalty0 1757--1759, 1983.
\newblock ISSN 0001-1452.
\newblock \doi{10.2514/3.8321}.
\newblock URL \url{http://arc.aiaa.org/doi/10.2514/3.8321}.

\bibitem[Boiko et~al.(1996)Boiko, Dovgal, Zanin, and Kozlov]{Boiko1996}
A.~V. Boiko, A.~V. Dovgal, B.~Yu. Zanin, and V.~V. Kozlov.
\newblock {Three-dimensional structure of separated flows on wings (review)}.
\newblock \emph{Thermophysics and Aeromechanics}, 3\penalty0 (1):\penalty0
  1--13, 1996.
\newblock ISSN 0869-8643.

\bibitem[Dell'Orso et~al.(2016)Dell'Orso, Tuna, and Amitay]{DellOrso2016a}
Haley Dell'Orso, Burak~A. Tuna, and Michael Amitay.
\newblock {Measurement of Three-Dimensional Stall Cells on a Two-Dimensional
  NACA0015 Airfoil}.
\newblock \emph{AIAA Journal}, 54\penalty0 (12):\penalty0 3872--3883, 2016.
\newblock ISSN 0001-1452.
\newblock \doi{10.2514/1.J054848}.
\newblock URL \url{http://arc.aiaa.org/doi/10.2514/1.J054848}.

\bibitem[Dell'Orso and Amitay(2018)]{DellOrso2018}
H.~Dell'Orso and M.~Amitay.
\newblock {Parametric investigation of stall cell formation on a NACA 0015
  airfoil}.
\newblock \emph{AIAA Journal}, 56\penalty0 (8):\penalty0 3216--3228, 2018.
\newblock ISSN 00011452.
\newblock \doi{10.2514/1.J056850}.

\bibitem[Neal and Amitay(2023)]{Neal2023}
J.~M. Neal and M.~Amitay.
\newblock {Three-dimensional separation over unswept cantilevered wings at a
  moderate Reynolds number}.
\newblock \emph{Physical Review Fluids}, 014703\penalty0 (8):\penalty0 1--24,
  2023.
\newblock \doi{10.1103/PhysRevFluids.8.014703}.

\bibitem[Yon and Katz(1998)]{Yon1998}
S.~A. Yon and J.~Katz.
\newblock {Study of the unsteady flow features on a stalled wing}.
\newblock \emph{AIAA Journal}, 36\penalty0 (3):\penalty0 305--312, 1998.
\newblock ISSN 00011452.
\newblock \doi{10.2514/2.372}.

\bibitem[McCullough and Gault(1951)]{McCullough1951}
G.~McCullough and D.~Gault.
\newblock {Examples of three representative types of airfoil-section stall at
  low speed}.
\newblock Technical report, Ames Aeronautical Laboratory, Moffett Field, CA,
  1951.

\bibitem[Broeren and Bragg(2001)]{Broeren2001}
A.~P. Broeren and M.~B. Bragg.
\newblock {Spanwise variation in the unsteady stalling flowfields of
  two-dimensional airfoil models}.
\newblock \emph{AIAA Journal}, 39\penalty0 (9):\penalty0 2001--1641, 2001.
\newblock ISSN 0001-1452.
\newblock \doi{10.2514/3.14912}.
\newblock URL \url{http://arc.aiaa.org/doi/abs/10.2514/3.14912}.

\bibitem[Manolesos and Voutsinas(2014)]{Manolesos2014}
Marinos Manolesos and Spyros~G. Voutsinas.
\newblock {Study of a stall cell using stereo particle image velocimetry}.
\newblock \emph{Physics of Fluids}, 26\penalty0 (4), 2014.
\newblock ISSN 10897666.
\newblock \doi{10.1063/1.4869726}.

\bibitem[Zhang et~al.(2020{\natexlab{a}})Zhang, Hayostek, Amitay, He,
  Theofilis, and Taira]{zhang2020formation}
K.~Zhang, S.~Hayostek, M.~Amitay, W.~He, V.~Theofilis, and K.~Taira.
\newblock On the formation of three-dimensional separated flows over wings
  under tip effects.
\newblock \emph{Journal of Fluid Mechanics}, 895, 2020{\natexlab{a}}.

\bibitem[Freymuth et~al.(1987)Freymuth, Finaish, and Bank]{freymuth1987further}
P.~Freymuth, F.~Finaish, and W~Bank.
\newblock Further visualization of combined wing tip and starting vortex
  systems.
\newblock \emph{AIAA journal}, 25\penalty0 (9):\penalty0 1153--1159, 1987.

\bibitem[Black(1956)]{Black1956}
J.~Black.
\newblock {Flow Studies of the Leading Edge Stall on a Swept-Back Wing at High
  Incidence}.
\newblock \emph{The Journal of the Royal Aeronautical Society}, 60\penalty0
  (541):\penalty0 51--60, 1956.
\newblock ISSN 0368-3931.
\newblock \doi{10.1017/s0368393100132390}.

\bibitem[Poll(1986)]{Poll1986}
D.~I. Poll.
\newblock {Spiral Vortex Flow Over a Swept-Back Wing.}
\newblock \emph{Aeronautical Journal}, 90\penalty0 (895):\penalty0 185--199,
  1986.
\newblock ISSN 00019240.

\bibitem[Yen and Huang(2009)]{Yen2009}
S.~C. Yen and L.~C. Huang.
\newblock {Flow patterns and aerodynamic performance of unswept and swept-back
  wings}.
\newblock \emph{Journal of Fluids Engineering, Transactions of the ASME},
  131\penalty0 (11):\penalty0 111101, 2009.
\newblock ISSN 00982202.
\newblock \doi{10.1115/1.4000260}.

\bibitem[Yen and Huang(2011)]{Yen2011}
S.~C. Yen and L.~C. Huang.
\newblock {Reynolds number effects on flow characteristics and aerodynamic
  performances of a swept-back wing}.
\newblock \emph{Aerospace Science and Technology}, 15\penalty0 (3):\penalty0
  155--164, 2011.
\newblock ISSN 12709638.
\newblock \doi{10.1016/j.ast.2010.10.005}.
\newblock URL \url{http://dx.doi.org/10.1016/j.ast.2010.10.005}.

\bibitem[Neal and Amitay(2022)]{neal_hire_swept}
J.~Neal and M.~Amitay.
\newblock {Three-Dimensional Flowfield Measurements of a Swept-back
  Cantilevered Wing at High Angles of Attack}.
\newblock In \emph{AIAA Paper 2022-0466}, 2022.

\bibitem[Zhang et~al.(2020{\natexlab{b}})Zhang, Hayostek, Amitay, Burtsev,
  Theofilis, and Taira]{zhang2020laminar}
K.~Zhang, S.~Hayostek, M.~Amitay, A.~Burtsev, V.~Theofilis, and K.~Taira.
\newblock Laminar separated flows over finite-aspect-ratio swept wings.
\newblock \emph{Journal of Fluid Mechanics}, 905:\penalty0 R1,
  2020{\natexlab{b}}.
\newblock \doi{10.1017/jfm.2020.778}.

\bibitem[Breitsamter and Laschka(2001)]{Breitsamter2001}
C.~Breitsamter and B.~Laschka.
\newblock {Vortical Flowield Structure at Forward Swept-Wing Configurations}.
\newblock \emph{Journal of Aircraft}, 38\penalty0 (2), 2001.

\bibitem[Zhang and Taira(2022)]{Zhang2022}
K.~Zhang and K.~Taira.
\newblock {Laminar vortex dynamics around forward-swept wings}.
\newblock \emph{Physical Review Fluids}, 7\penalty0 (2):\penalty0 1--11, 2022.
\newblock ISSN 2469990X.
\newblock \doi{10.1103/PhysRevFluids.7.024704}.

\bibitem[Skinner et~al.(2020)Skinner, Green, and Zare-Behtash]{Skinner2020}
S.~N. Skinner, R.~B. Green, and H.~Zare-Behtash.
\newblock {Wingtip vortex structure in the near-field of swept-tapered wings}.
\newblock \emph{Physics of Fluids}, 32\penalty0 (9), 2020.
\newblock ISSN 10897666.
\newblock \doi{10.1063/5.0016353}.
\newblock URL \url{https://doi.org/10.1063/5.0016353}.

\bibitem[Zhang et~al.(2019)Zhang, Jaworski, McParlin, and
  Turner]{zhang_jaworski_mcparlin_turner_2019}
S.~Zhang, A.~J. Jaworski, S.~C. McParlin, and J.~T. Turner.
\newblock Experimental investigation of the flow structures over a 40° swept
  wing.
\newblock \emph{The Aeronautical Journal}, 123\penalty0 (1259):\penalty0
  39–55, 2019.
\newblock \doi{10.1017/aer.2018.118}.

\bibitem[Ribeiro et~al.(2023{\natexlab{b}})Ribeiro, Taira, Neal, Amitay,
  Burtsev, and Theofilis]{Ribeiro:AIAA23}
J.~H.~M. Ribeiro, K.~Taira, J.~M. Neal, M.~Amitay, A.~Burtsev, and
  V.~Theofilis.
\newblock Wake dynamics of tapered wings. part i: a computational study.
\newblock In \emph{AIAA Paper 2023-2297}, 2023{\natexlab{b}}.

\bibitem[Neal et~al.(2023)Neal, Amitay, Burtsev, Theofilis, Ribeiro, and
  Taira]{Neal:AIAA23}
J.~M. Neal, M.~Amitay, A.~Burtsev, V.~Theofilis, J.~H.~M. Ribeiro, and
  K.~Taira.
\newblock Wake dynamics of tapered wings. part ii: an experimental study.
\newblock In \emph{AIAA Paper 2023-2298}, 2023.

\bibitem[Burtsev et~al.(2023)Burtsev, Theofilis, Ribeiro, Taira, Neal, and
  Amitay]{Burtsev:AIAA23}
A.~Burtsev, V.~Theofilis, J.~H.~M. Ribeiro, K.~Taira, J.~M. Neal, and
  M.~Amitay.
\newblock Wake dynamics of tapered wings. part iii: tri{G}lobal linear
  stability analysis.
\newblock In \emph{AIAA Paper 2023-2299}, 2023.

\bibitem[Hayostek et~al.(2019)Hayostek, Amitay, Zhang, Taira, He, and
  Theofilis]{hayoss19}
S.~Hayostek, M.~Amitay, K.~Zhang, K.~Taira, W.~He, and V.~Theofilis.
\newblock Wake dynamics of finite aspect ratio wings. part i: An experimental
  study.
\newblock In \emph{AIAA Paper 2019-1384}, 2019.

\end{thebibliography}

\end{document}